 \documentclass[seceq]{ptptex}

 \usepackage{graphicx}

\newcommand{\lax}{\>\vcenter{\hbox{$<$\hskip-.75em\lower1.0ex\hbox{$\sim$}}}\>}
\newcommand{\gax}{\>\vcenter{\hbox{$>$\hskip-.75em\lower1.0ex\hbox{$\sim$}}}\>}
\newcommand{\ion}[2]{#1$\,${\small{#2}}\relax}
\newcommand{\Mdot}{\dot M}
\newcommand{\Msun}{\rm M_\odot}

\newcommand{\yr}{\rm yr^{-1}}

\title{
Hydrodynamic and Spectral Simulations of HMXB Winds
}

\author{
Christopher W.~{\scshape Mauche}$^{1,}$\footnote{E-mail: 
mauche@cygnus.llnl.gov},
Duane A.~{\scshape Liedahl}$^{1,}$\footnote{E-mail: liedahl1@llnl.gov},
Shizuka {\sc Akiyama}$^{1,}$\footnote{E-mail: shizuka@llnl.gov},
and Tomek {\scshape Plewa}$^{2,}$\footnote{E-Mail: tomek@uchicago.edu}
}

\inst{
$^1$Lawrence Livermore National Laboratory, Livermore, CA 94550, USA \\
$^2$University of Chicago, 5640 South Ellis Avenue, RI-475, Chicago, IL
     60615, USA
}

\abst{
We describe preliminary results of a global model of the radiatively-driven
photoionized wind and accretion flow of the high-mass X-ray binary Vela X-1.
The full model combines FLASH hydrodynamic calculations, XSTAR photoionization
calculations, HULLAC atomic data, and Monte Carlo radiation transport. We
present maps of the density, temperature, velocity, and ionization parameter
from a FLASH two-dimensional time-dependent simulation of Vela X-1, as well
as maps of the emissivity distributions of the X-ray emission lines. }

\begin{document}

\maketitle

\section{Introduction}

As described by Castor, Abbott, and Klein (hereafter CAK),\cite{rf:cak75}
mass loss in the form of a high velocity wind is driven from the surface of
an OB star by radiation pressure on a multitude of resonance transitions of
intermediate charge states of cosmically abundant elements. The wind is
characterized by a mass-loss rate $\Mdot \sim 10^{-6}$--$10^{-5}~\Msun~\yr $
and a velocity profile $V(R) \sim V_\infty (1-R_{\rm OB} /R)^\beta$, where
$\beta\approx\frac{1}{2}$, the terminal velocity $V_\infty\sim 3\,V_{\rm esc}
= 3\,(2GM_{\rm OB}/R_{\rm OB})^{1/2} \sim 1500~\rm km~s^{-1}$, $R$ is the
distance from the OB star, and $M_{\rm OB}$ and $R_{\rm OB}$ are respectively
the mass and radius of the OB star. In a detached high-mass X-ray binary
(HMXB), a compact object, typically a neutron star, captures a fraction
$f\sim\pi R_{\rm BH}^2 /4\pi a^2$ of the OB star wind, where $a$ is the binary
separation, $R_{\rm BH} = 2\, GM_{\rm NS} /[V(a)^2+c_{\rm s}^2]$ is the
Bondi-Hoyle radius, $c_{\rm s} \sim 10\, (T/10^4)^{1/2}~\rm km~s^{-1}$ is the
sound speed, and $T$ is the wind temperature. Accretion of this material onto
the neutron star powers an X-ray luminosity $L_{\rm X} \sim fG\Mdot M_{\rm NS}
/R_{\rm NS}\sim 10^{36}$--$10^{37}~\rm erg~s^{-1}$, where $M_{\rm NS}$ and
$R_{\rm NS}$ are respectively the mass and radius of the neutron star. The
resulting X-ray flux photoionizes the wind and reduces its ability to be
radiatively driven, both because the higher ionization state of the plasma
results is a reduction in the number of resonance transitions, and because the
energy of the transitions shifts to shorter wavelengths where the overlap with
the stellar continuum is lower. To first order, the lower radiative driving
results in a reduced wind velocity near the neutron star $V(a)$, which
increases the Bondi-Hoyle radius $R_{\rm BH}$, which increases the accretion
efficiency $f$, which increases the X-ray luminosity $L_{\rm X}$. In this way,
the X-ray emission of HMXBs is the result of a complex interplay between the
radiative driving of the wind of the OB star and the photoionization of the
wind by the neutron star.

Known since the early days of X-ray astronomy, HMXBs have been extensively
studied observationally, theoretically,\cite{rf:hat77,rf:mcc84,rf:ste90} and
computationally.\cite{rf:blo90,rf:blo91,rf:blo94,rf:blo95} They are excellent
targets for X-ray spectroscopic observations because the large covering
fraction of the wind and the moderate X-ray luminosities result in large
volumes of photoionized plasma that produce strong recombination lines and
narrow radiative recombination continua of H- and He-like ions, as well as
fluorescent lines from lower charge\vspace{-0.25\baselineskip} states.

\section{Vela\vspace{-0.25\baselineskip} X-1}

Vela X-1 is the prototypical detached HMXB, having been studied extensively
in nearly every waveband, particularly in X-rays, since its discovery as an
X-ray source during a rocket flight four decades ago. It consists of a B0.5
Ib supergiant and a magnetic neutron star in an 8.964-day orbit. From an
X-ray spectroscopic point of view, Vela X-1 distinguished itself in 1994 when
Nagase et al.,\cite{rf:nag94} using {\it ASCA\/} SIS data, showed that, in
addition to the well-known 6.4 keV emission line, the eclipse X-ray spectrum
is dominated by recombination lines and continua of H- and He-like Ne, Mg,
Si, S, Ar, and Fe. These data were subsequently modeled in detail by Sako et
al.,\cite{rf:sak99} using a  kinematic model in which the photoionized wind
was characterized by the ionization parameter $\xi\equiv L_{\rm X}/nr^2$,
where $r$ is the distance from the neutron star and $n$ is the number
density, given by the mass-loss rate and velocity law of an undisturbed
CAK wind. Vela X-1 was subsequently observed with the {\it Chandra\/} HETG
in 2000 for 30 ks in eclipse\cite{rf:sch02} and in 2001 for 85, 30, and 30
ks in eclipse and at binary phases 0.25 and 0.5,
respectively.\cite{rf:gol04,rf:wat06} Watanabe et al.\cite{rf:wat06}, using
very similar assumptions as Sako et al. and a Monte Carlo radiation transfer
code, produced a global model of Vela X-1 that simultaneously fit the HETG
spectra from the three binary phases with a wind mass-loss rate $\Mdot\approx
2\times 10^{-6}~\rm\Msun~\yr $ and terminal velocity $V_\infty = 1100~\rm
km~s^{-1}$. One of the failures of this model was the velocity shifts of the
emission lines between eclipse and phase 0.5, which were observed to be
$\Delta V\approx 400$--$500~\rm km~s^{-1}$, while the model simulations
predicted $\Delta V\sim 1000~\rm km~s^{-1}$. In order to resolve this
discrepancy, Watanabe et al.\ performed a 1D calculation to estimate the wind
velocity profile along the line of centers between the two stars, accounting,
in an approximate way, for the reduction of the radiative driving due to
photoionization. They found that the velocity of the wind near the neutron
star is lower by a factor of 2--3 relative to an undisturbed CAK wind, which
was sufficient to explain the observations. However, these results were not
fed back into their global model to determine the effect on the X-ray
\vspace{-0.25\baselineskip} spectra.

\section{Hydrodynamic\vspace{-0.25\baselineskip} Simulations}

To make additional progress in our understanding of the wind and accretion
flow of Vela X-1 in particular and HMXBs in general --- to bridge the gap
between the detailed hydrodynamic models of Blondin et al.\ and the simple
kinetic-spectral models of Sako et al.\ and Watanabe et al.\ --- we have
undertaken a project to develop improved models of radiatively-driven
photoionized accretion flows, with the goal of producing synthetic X-ray
spectral models that possess a level of detail commensurate with the grating
spectra returned by {\it Chandra\/} and {\it XMM-Newton\/}.

\begin{figure}
\centerline{\includegraphics[height=97pt]{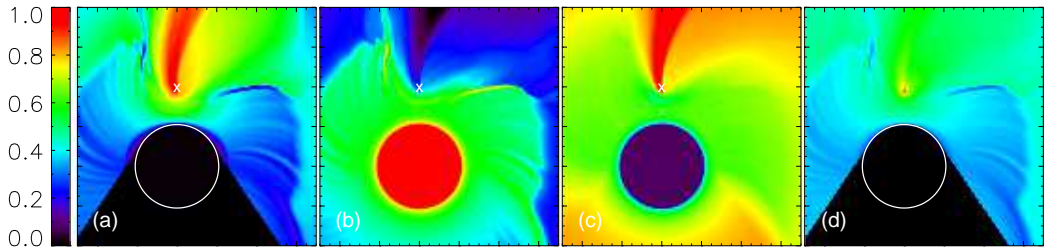}}
\vspace{-7pt}
\caption{Color-coded maps of
({\it a\/}) $\log T\,  ({\rm K})=[4.4,8.3]$,
({\it b\/}) $\log n\,  ({\rm cm}^{-3})=[7.4,10.8]$,
({\it c\/}) $\log V\,  ({\rm km~s^{-1}})=[1.3,3.5]$, and
({\it d\/}) $\log\xi\, ({\rm erg~cm~s^{-1}})=[1.1,7.7]$ in the orbital plane
of Vela~X-1. The positions of the OB star and neutron star are shown by the
circle and the ``$\times $,'' respectively. The horizontal axis $x=[-5,7]
\times 10^{12}$ cm, and the vertical axis $y=[-4,8]\times 10^{12}$ cm.}
\label{fig:1}
\vspace{-10pt}
\end{figure}

This project combines (1) XSTAR\cite{rf:kal01} photoionization calculations,
(2) HULLAC\cite{rf:bar88} emission models appropriate to X-ray photoionized
plasmas, (3) improved models of the radiative driving of the photoionized
wind, (4) FLASH\cite{rf:fry00} three-dimensional time-dependent adaptive-mesh
hydrodynamics calculations, and (5) a Monte Carlo radiation transport
code.\cite{rf:mau04} Radiative driving of the wind is accounted for via the
force multiplier formalism,\cite{rf:cak75} accounting for X-ray
photoionization and non-LTE population kinetics using HULLAC atomic data for
$2\times 10^6$ lines of $35{,}000$ energy levels of 166 ions of the 13 most
cosmically abundant elements. In addition to the usual hydrodynamic quantities,
the FLASH calculations account for
({\it a\/}) the gravity of the OB star and neutron star,
({\it b\/}) Coriolis and centrifugal forces,
({\it c\/}) radiative driving of the wind as a function of the local
             ionization parameter, temperature, and optical depth,
({\it d\/}) photoionization and Compton heating of the irradiated wind, and
({\it e\/}) radiative cooling of the irradiated wind and the ``shadow wind''
             behind the OB star.
To demonstrate typical results of our simulations, we show in Fig.~1
color-coded maps of the log of the ({\it a\/}) temperature, ({\it b\/})
density, ({\it c\/}) velocity, and ({\it d\/}) ionization parameter of a FLASH
simulation with parameters appropriate to Vela X-1. This is a 2D simulation
in the binary orbital plane, has a resolution of $\Delta l=9.4\times
10^{10}$ cm, and, at the time step shown ($t = 100$ ks), the relatively slow
($V\approx 400~\rm km~s^{-1}$)\footnote{Note that this velocity reproduces
the value that Watanabe et al.\ found was needed to match the velocity of the
emission lines in the {\it Chandra\/} HETG spectra of Vela X-1.} irradiated
wind has reached just $\sim 2$ stellar radii from the stellar surface. The
various panels show (1) the effect of the Coriolis and centrifugal forces,
which cause the flow to curve clockwise, (2) the cool, fast wind behind the
OB star, (3) the hot, slow irradiated wind, (4) the hot, low density, high
velocity flow downstream of the neutron star, and (5) the bow shock and two
flanking shocks formed where the irradiated wind collides with the hot
disturbed flow in front and downstream of the neutron star.

\begin{figure}
\centerline{\includegraphics[height=97pt]{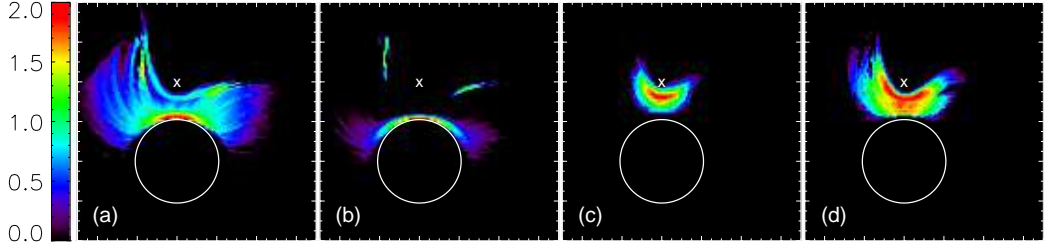}}
\vspace{-7pt}
\caption{Color-coded maps of the log of the X-ray emissivity of
({\it a\/}) \ion{Si}{XIV}  Ly$\alpha $,
({\it b\/}) \ion{Si}{XIII} He$\alpha $,
({\it c\/}) \ion{Fe}{XXVI} Ly$\alpha $, and
({\it d\/}) \ion{Fe}{XXV}  He$\alpha $.
In each case, two orders of magnitude are plotted.}
\label{fig:2}
\vspace{-10pt}
\end{figure}

Given these maps, it is straightforward to determine where in the binary the
X-ray emission originates. To demonstrate this, we show in Fig.~2
color-coded maps of the log of the emissivity of
({\it a\/}) \ion{Si}{XIV}  Ly$\alpha $,
({\it b\/}) \ion{Si}{XIII} He$\alpha $,
({\it c\/}) \ion{Fe}{XXVI} Ly$\alpha $, and
({\it d\/}) \ion{Fe}{XXV}  He$\alpha $.
The gross properties of these maps agree with Fig.~24 of Watanabe et al., but
they are now (1) quantitative rather than qualitative and (2) specific to
individual transitions of individual ions. The maps also capture features that
otherwise would not have been supposed, such as the excess emission in the H-
and He-like Si lines downstream of the flanking shocks. Combining these maps
with the velocity map (Fig.~1{\it c\/}), these models make very specific
predictions about (1) the intensity of the emission features, (2) where the
emission features originate, and (3) their velocity widths and amplitudes as
a function of binary phase.

The next step in our modeling effort is to feed the output of the FLASH
simulations into the Monte Carlo radiation transfer code, to determine how the
spatial and spectral properties of the X-ray emission features are modified by
Compton scattering, photoabsorption followed by radiative cascades, and line
scattering. This work is underway.

\section*{Acknowledgements}

This work was performed under the auspices of the U.S. Department of Energy
by University of California, Lawrence Livermore National Laboratory under
Contract W-7405-Eng-48. T.~Plewa's contribution to this work was supported
in part by the U.S.\ Department of Energy under Grant No.\ B523820 to the
Center for Astrophysical Thermonuclear Flashes at the University of Chicago.


\end{document}